\documentclass[preprint,onecolumn,superscriptaddress,tightenlines,nofootinbib,12pt]{article}
\pdfoutput=1
\usepackage{graphicx}
\usepackage{epsfig}
\usepackage{amsmath,latexsym,amssymb,hyperref,enumerate}
\usepackage{amsfonts}
\usepackage{url}
\usepackage{color}
\usepackage{cancel}
\usepackage{cite}
\hypersetup{colorlinks,citecolor= nicegreen,linkcolor= nicered}
\definecolor{nicered}{rgb}{0.7,0.1,0.1}
\definecolor{nicegreen}{rgb}{0.1,0.5,0.1}
\newcommand{\bg}[1]{\mbox{\boldmath{$#1$}}}

\begin{document}
\renewcommand{\thefootnote}{\fnsymbol{footnote}}
\renewcommand{\baselinestretch}{1.2}
\normalsize
\vspace{-0.5cm}
\begin{flushright}
OSU-HEP-11-08\\
November 21, 2011 \\
\vspace{1cm}
\end{flushright}

\vspace*{-0.5cm}
\centerline{\bf\Large Perturbative unitarity constraints on general $W'$ models}
\vspace{2mm}
\centerline{\bf \Large and collider implications}
\vspace*{1.cm}

\centerline{\bf K.S. Babu\footnote{Email:
babu@okstate.edu}, ~J. Julio\footnote{Email:
julio.julio@okstate.edu}}

\centerline{\em Department of Physics and Oklahoma Center for High Energy Physics,}
\vspace*{-0.5mm}
\centerline{\em Oklahoma State University, Stillwater, OK 74078, USA}

\vspace*{0.5cm}

\centerline{\bf Yue Zhang\footnote{Email: yuezhang@ictp.it}}
\centerline{\em Abdus Salam International Centre for Theoretical Physics,}
\centerline{\em Strada Costiera 11, Trieste 34014, Italy}

\vspace{0.5cm}

\begin{abstract}

We study perturbative unitarity constraints on general $W'$ models by considering the high energy behavior
of fermion scattering into gauge bosons. In most cases we survey, a $Z'$ boson with a comparable mass must be present
for the theory to be consistent, with fixed couplings to the standard model gauge bosons and fermions. Applying these
results to a class of $W'$ models which explains the top quark forward--backward asymmetry
observed at the Tevatron, we find that a $Z'$ must exist with a mass below 7$-$8\,TeV and sizable coupling
to the light quarks.  While such a $Z'$ is strongly constrained by existing experiments,
we show that the LHC can explore the entire mass range up to the unitarity limit. We also show how it is possible, by raising the
$Z'$ mass consistent with unitarity, to explain the CDF $Wjj$ excess in terms of a light $W'$, without generating
an excess in $Zjj$ events.

\end{abstract}


\renewcommand{\thefootnote}{\arabic{footnote}}

\newpage
\section{Introduction}

It is a well established fact that in a weakly coupled renormalizable theory, the high-energy behavior of scattering amplitudes should not violate unitarity under perturbative calculations~\cite{mathur,Cornwall:1974km}. In the standard model (SM), an upper
limit on the Higgs boson mass of about a TeV has been inferred based on unitarity in the gauge boson ($V$) scattering process
$VV\to VV$ \cite{Lee:1977yc, Lee:1977eg}.
Prior to the discovery of the $Z$ boson, it was shown, from the unitarity in the high energy scattering of various types
that the coupling of the $Z$ boson to the SM fermions must be unique~\cite{Cornwall:1973tb}. Perturbative unitarity has also proved to be a useful
guide for studying beyond the SM physics, often in a model--independent fashion.

A charged gauge bosons, $W'$, arises in a variety of extensions of the SM.
It is present in theoretical frameworks such as left-right symmetric models -- as the right--handed parity partner of the $W$ gauge boson~\cite{Mohapatra:1974gc}, in Kaluza-Klein theories of extra dimensions -- as the excitation of the $W$~\cite{KK,Agashe:2003zs}, in little Higgs theories -- as the gauge bosons of the extended symmetry~\cite{ArkaniHamed:2001ca}, and in several other well-motivated extensions of the SM. With the running of the Large Hadron Collider (LHC), $W'$ models have drawn renewed interest~\cite{Grojean:2011vu} as it is on the list of particles subject to direct searches. Furthermore, recently there have been new phenomenological motivations for having a light charged gauge boson. A $W'$ boson with mass below a TeV  and coupling to top and down quarks has been introduced as one of the leading explanations \cite{Barger:2010mw, Cao:2010zb, Gresham:2011pa, Jung:2011zv} for the anomalous top-quark forward-backward asymmetry observed at the Tevaron~\cite{:2007qb,Aaltonen:2011kc}. There are also proposals in the literature which utilize a light $W'$ with a mass of order 150 GeV to explain the $W$ plus dijet event excess~\cite{Wang:2011uq} observed by the CDF collaboration~\cite{Aaltonen:2011mk}.

In this paper we study the theoretical and phenomenological constraints on $W'$ models assuming that it is the remnant of spontaneous symmetry breaking through the Higgs mechanism. We take a model-independent approach in the sense that the mass of $W'$ and its coupling to SM fermions are the only input information that will be used. We study the implication of perturbative unitarity on this setup from two body scattering of fermions into gauge boson final states. Throughout this paper, we consider weakly coupled theories, since we focus on the weak interaction sector where
there is no evidence for strong dynamics.  Violation of perturbative unitarity in low-energy effective theory does not necessarily imply inconsistency
in a strongly coupled theory, but would rather indicate appearance of resonances or other such non-perturbative objects.

Our main conclusion is that generically a $Z'$ gauge boson should also be present along with the $W'$ in a consistent theory. Unitarity of the theory fixes the couplings of the $Z'$ to SM gauge bosons and fermions, and implies an upper bound on its mass.
We derive general formulas that relate the various couplings and masses and apply them
to specific models with purely left-handed or purely right-handed $W'$ couplings to the fermions. 
In the class of $W'$ models that have been suggested for $t\bar t$ asymmetry, a light (sub-TeV) $W'$
has been postulated with flavor-changing coupling of the type $W'\bar td$. We point out that the same coupling also leads to the scattering of a pair of fermions (e.g., $d \bar{d}$) into $W'^+W'^-$. Although such $W'$ could evade all current constraints with the particular choice of quark flavors, the $Z'$ boson is subject to more severe direct search limits. Unitarity sets an upper bound on its mass which we find to be less than about 7$-$8\,TeV.  The LHC is capable of probing such a $Z'$ for most part of this mass window, as we show. We also note that unitarity restoration would imply significant destructive interference when each diagram contributing to a given process individually has a bad high energy behavior.
Therefore, a large hierarchy between $Wjj$ and $Zjj$ cross sections can be achieved by consistently
raising the mass of $Z'$ which unitarizes the $f \overline{f} \rightarrow W'^+ W'^-$ process.
This could be interesting for $W'$ models explaining the CDF $Wjj$ event excess.
For completeness, we also classify realistic models that contain the SM gauge symmetry and has a charged $W'$ gauge boson (in a more
general sense) but no $Z'$ gauge boson, and show how unitarity is preserved in this class of models.

\section{General framework}

\begin{figure}[t!]
\begin{center}
\includegraphics[width=4cm]{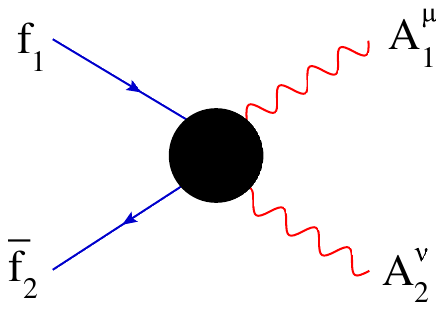}
\caption{A pair of fermions scattering into a pair of gauge bosons.}\label{general}
\end{center}
\end{figure}

We will focus on the class of processes in which a pair of fermions scatter into a pair of gauge bosons, $f_1 \bar f_2 \to A^\mu_1 A^\nu_2$, and study its unitary properties at very high center-of-mass energy. As pointed out in~Ref. \cite{Cornwall:1974km},
the asymptotic form of a four-particle amplitude $\mathcal{A}$ should not grow with energy.
The general process is shown in Fig.~\ref{general}. In the following we will consider tree-level process only. At very high energies,
in a spontaneously broken gauge theory, due to the Goldstone boson equivalence theorem,
the contributions to the scattering amplitude is dominated by
the longitudinal components of the final state gauge bosons.

\begin{figure}[t!]
\begin{center}
\includegraphics[width=3.5cm]{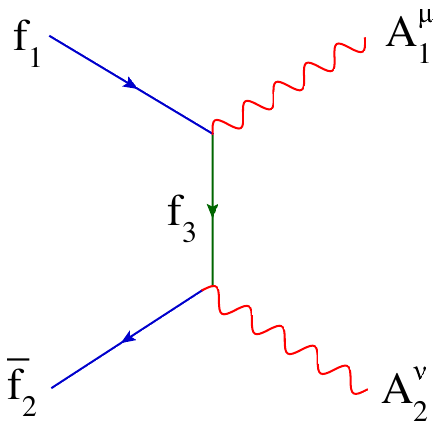}\hspace{0.5cm}
\includegraphics[width=3.5cm]{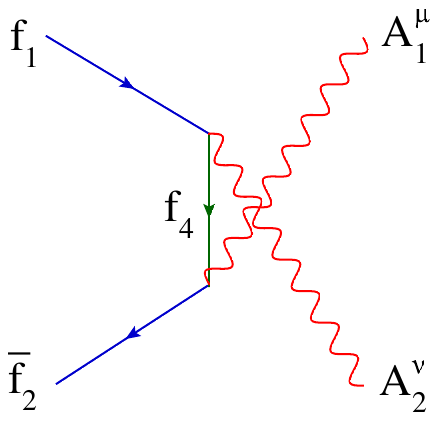}\hspace{0.5cm}
\includegraphics[width=3.5cm]{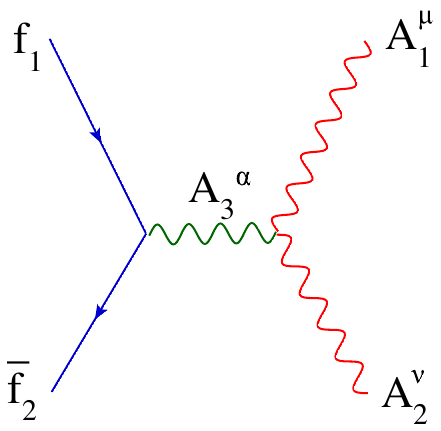}\hspace{0.5cm}
\includegraphics[width=3.5cm]{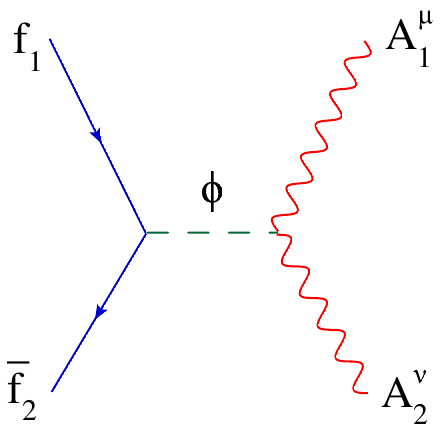}
\caption{Different topologies of the scattering  $f_1 \bar f_2 \to A^\mu_1 A^\nu_2$. All the internal particles
must have masses not far above $M_{A_1, A_2}$, otherwise their effects
will decouple.}\label{specific}
\end{center}
\end{figure}

At the tree--level, the processes above can happen through the following channels: a) t-channel fermion $f_3$ exchange, b) u-channel fermion $f_4$ exchange, c) s-channel gauge boson $A_{3}$ exchange, and d) s-channel scalar $\varphi$ exchange. The corresponding topologies are shown in Fig.~\ref{specific}.

We define the relevant couplings which will be used in our analysis.
The couplings between gauge boson $A_i^\mu$ with two fermions $f_j$, $\bar f_k$ can be written in the general form as
\begin{eqnarray}
\mathcal{L} = g_{A_i f_j \bar f_k}^{L,R} \bar f_k \gamma^\mu P_{L,R} f_j A_{i\mu} + {\rm h.c.} \
\end{eqnarray}
where $P_{L,R} = (1\mp \gamma_5)/2$ are the chirality projection operators.
The coupling between three gauge bosons $A_1$, $A_2$ and $A_3$ can be written in the general gauge and Lorentz invariant form as
\begin{eqnarray}
\mathcal{L}_{A_1 A_2 A_3} &=& i g_{A_1 A_2 A_3} \hat{\mathcal{L}}_{A_1 A_2 A_3} \ ,
\end{eqnarray}
with
\begin{eqnarray}
\hat{\mathcal{L}}_{A_1 A_2 A_3} &=&  (\partial_\mu A_{1\nu})(A_2^\mu A_3^\nu - A_2^\nu A_3^\mu) - (\partial_\mu A_{2\nu})(A_1^\mu A_3^\nu - A_1^\nu A_3^\mu) + (\partial_\mu A_{3\nu})(A_1^\mu A_2^\nu - A_1^\nu A_2^\mu) \ .
\label{gauge-trilinear}
\end{eqnarray}
The coupling between a scalar field $\varphi$ with fermions $f_j$, $\bar f_k$ can be written as
\begin{eqnarray}
\mathcal{L} = y_{\varphi f_j \bar f_k}^{LR,RL} \bar f_k P_{L,R} f_j \varphi + {\rm h.c.} \ .
\end{eqnarray}
The coupling between a scalar field $\varphi$ with two gauge bosons $A_1^\mu$ and $A_2^\nu$ is written as
\begin{eqnarray}
\mathcal{L} = v_\varphi g_{A_1 A_2 \varphi} \varphi A_1^\mu A_{2\mu} \ .
\end{eqnarray}

Perturbative unitarity constraints are twofold. First, the coherent sum of all relevant amplitudes should be finite when the energy scale goes to infinity. Second, in the partial wave expansion~\cite{Chanowitz:1978mv}, each finite partial wave of the amplitude should satisfy $|a^J_{\lambda}|<1$, where
\begin{eqnarray}\label{partial}
a^J_\lambda = \frac{1}{32\pi} \int_{-1}^{1} d (\cos\theta) d_{\lambda 0}^J(\theta) \mathcal{A}_{\lambda} \ ,
\end{eqnarray}
and where $\lambda=(\lambda_{f_1} -\lambda_{\bar f_2})/2$ is the helicity difference between two initial state fermions. $d_{\lambda \lambda'}^J(\theta)$ are the Wigner $d$-functions~\cite{Jacob:1959at}. The leading functions relevant for our study are
\begin{eqnarray}
d_{0 0}^0(\theta) = 1, \ \ d_{1 0}^1(\theta) = -\frac{\sin\theta}{\sqrt{2}} \ .
\end{eqnarray}
The first requirement implies sum rules among different couplings involved in the scattering process, while the second one leads to constraints on the masses of the states being exchanged in different channels.

According to the initial state fermion helicity, the scattering can be divided into two classes.
In the first class, the helicity of initial state fermions $f_1$ is opposite to that of $\bar f_2$ (or the same as $f_2$), which implies even number of fermion mass insertions in the process. We call this the helicity violating case, since the initial state has nonzero total helicity $(\lambda=1)$ while the final state has $\lambda=0$. Such processes can involve t- and u-channels fermion exchange as well as s-channel gauge boson exchange.
In this case, the amplitude can be expanded in terms of the center-of-mass energy squared $s$ as
\begin{eqnarray}
\mathcal{A}^{(-+)}_{f_{1}\bar f_{2}\to A_1 A_2} &=& \mathcal{A}_{1} s + \mathcal{A}_{0} + \mathcal{A}_{-1} s^{-1} + \mathcal{A}_{-2} s^{-2} + \cdots \ .
\end{eqnarray}
The finiteness of $\mathcal{A}^{(-+)}_{f_1\bar f_2\to A_1 A_2}$ implies $\mathcal{A}_1=0$ and therefore a sum rule among the relevant couplings:
\begin{eqnarray}\label{sumrule1}
&&\sum_{f_3} g_{A_1 f_1 \bar f_3}^{L} g_{A_2 f_3 \bar f_2}^{L} - \sum_{f_4} g_{A_1 f_4 \bar f_2}^{L} g_{A_2 f_1 \bar f_4}^{L} = \sum_{A_3} g^L_{A_3 f_1 \bar f_2} g_{A_1A_2A_3}  \ ,
\end{eqnarray}
There is a similar sum rule with $L\leftrightarrow R$ in the couplings in Eq. (\ref{sumrule1}), which arises from
 $\mathcal{A}^{(+-)}_{f_1\bar f_2\to A_1 A_2}$.
If these conditions are not satisfied, the partial wave amplitude would increase with energy, and one should expect new states to appear
with the proper couplings around the scale $\Lambda$ with
\begin{eqnarray}
\Lambda = {\rm Min} \left\{\sqrt{\frac{32\pi} {\int_{-1}^{1} d (\cos\theta) d_{\lambda 0}^J(\theta) \mathcal{A}_{1, \lambda}}}, \ \ {\rm for\ all\ }J\right\} \ .
\end{eqnarray}
When the rum rule Eq. (\ref{sumrule1}) is satisfied, $\mathcal{A}_1=0$, and the theory can remain perturbative.
The finite amplitude at very high energy can be written as
\begin{footnotesize}
\begin{eqnarray}\label{finite}
\mathcal{A}_0 &=&\ - \sum_{f_3} g_{A_1 f_1 \bar f_3}^{L} g_{A_2 f_3 \bar f_2}^{L}\frac{M_{A_1}^2 + M_{A_2}^2 - \frac{1}{2} (m_{f_1}^2 + m_{f_2}^2) - m_{f_3}^2 \csc^2 \frac{\theta}{2} }{2M_{A_1} M_{A_2}} \sin\theta \nonumber \\
&&\ + \sum_{f_3} g_{A_1 f_1 \bar f_3}^{R} g_{A_2 f_3 \bar f_2}^{R} \frac{m_{f_1} m_{f_2}\cot^2\frac{\theta}{2}}{2M_{A_1} M_{A_2}}\sin\theta - \sum_{f_3} \left( g_{A_1 f_1 \bar f_3}^{L} g_{A_2 f_3 \bar f_2}^{R} m_{f_2} + g_{A_1 f_1 \bar f_3}^{R} g_{A_2 f_3 \bar f_2}^{L}  m_{f_1} \right) \frac{m_{f_3}\csc^2\frac{\theta}{2}}{2M_{A_1} M_{A_2}} \sin\theta \nonumber \\
&&\ + \sum_{f_4} g_{A_1 f_4 \bar f_2}^{L} g_{A_2 f_1 \bar f_4}^{L} \frac{M_{A_1}^2 + M_{A_2}^2 - \frac{1}{2} (m_{f_1}^2 + m_{f_2}^2) - m_{f_4}^2 \sec^2 \frac{\theta}{2}}{2M_{A_1} M_{A_2}} \sin\theta \nonumber \\
&&\ - \sum_{f_3} g_{A_1 f_1 \bar f_3}^{R} g_{A_2 f_3 \bar f_2}^{R} \frac{m_{f_1} m_{f_2}\tan^2\frac{\theta}{2}}{2M_{A_1} M_{A_2}}\sin\theta + \sum_{f_3} \left( g_{A_1 f_1 \bar f_3}^{L} g_{A_2 f_3 \bar f_2}^{R} m_{f_1} + g_{A_1 f_1 \bar f_3}^{R} g_{A_2 f_3 \bar f_2}^{L} m_{f_2} \right) \frac{m_{f_3}\sec^2\frac{\theta}{2}}{2M_{A_1} M_{A_2}} \sin\theta \nonumber \\
&&\ + \sum_{A_3} g^L_{A_3 f_1 \bar f_2} g_{A_1A_2A_3} \frac{  M_{A_3}^2 - \frac{1}{2} (m_{f_1}^2 + m_{f_2}^2) }{2M_{A_1} M_{A_2}} \sin\theta + \sum_{A_3} g^R_{A_3 f_1 \bar f_2} g_{A_1A_2A_3} \frac{  m_{f_1} m_{f_2} }{2M_{A_1} M_{A_2}} \sin\theta, \ \ \ (\theta\neq 0, \pi) \ .
\end{eqnarray}
\end{footnotesize}
\noindent A similar expression for the finite amplitude holds for $\mathcal{A}^{(+-)}_{f_1\bar f_2\to A_1 A_2}$ with
the replacement $L\leftrightarrow R$ in in the couplings
in Eq. (\ref{finite}). Demanding the partial waves of $A_0$ to satisfy unitarity, one can obtain an upper bound on the masses of 
the various particles involved in the scattering process.

The singularities at $\theta=0, \pi$ in Eq. (\ref{finite}) are not physical. They simply reflect the breakdown of the expansion in term of $\frac{m_t^2}{s (1\pm \cos\theta)}$ in the t-channel amplitude. The $\theta$ integral in Eq.~(\ref{partial}) can be performed $-1+\epsilon$ to $1-\epsilon$. Here $\epsilon$ can be taken infinitesimal because for any $\theta$, one can always find sufficiently large $s$ such that the above expansion is valid. In fact, all of the partial wave amplitudes above are finite, so are the total cross sections. As we noted already, the total helicities of initial and final states differ by one unit. Therefore, in the massless fermion limit the scattering must happen in the p-wave, i.e., there is an additional angular factor $\sin\theta$ which removes any singularity in the partial wave expansion.

The second class of scattering involves $f_1$  and $\bar f_2$ ($f_2$) with the same (opposite) helicity.  We call it the helicity conserving case because both initial and final states have vanishing total helicity. In this case, there must be odd number of mass insertions along the fermion line. In this helicity conserving case, the amplitude can be expanded in terms of ``half-integer" powers of $s$,
\begin{eqnarray}\label{massinsertion}
\mathcal{A}^{(--)}_{f_1\bar f_2\to A_1 A_2} &=& \mathcal{A}_{\frac{1}{2}} s^{\frac{1}{2}} + \mathcal{A}_{-\frac{1}{2}} s^{-\frac{1}{2}} + \mathcal{A}_{-\frac{3}{2}} s^{-\frac{3}{2}} + \cdots \ .
\end{eqnarray}
The Feynman diagrams generally include t- and u-channels fermion exchange and s-channel scalar boson exchange and vanishing $\mathcal{A}_{\frac{1}{2}}$ implies a sum rule
\begin{footnotesize}
\begin{eqnarray}\label{sumrule2}
&&\sum_{f_3} m_{f_3} g_{A_1 f_1 \bar f_3}^{L} g_{A_2 f_3 \bar f_2}^{R} - \sum_{f_3} m_{f_1} g_{A_1 f_1 \bar f_3}^{R} g_{A_2 f_3 \bar f_2}^{R} \cos^2\frac{\theta}{2}  - \sum_{f_3} m_{f_2} g_{A_1 f_1 \bar f_3}^{L} g_{A_2 f_3 \bar f_2}^{L} \cos^2\frac{\theta}{2} \nonumber \\
&& + \sum_{f_4} m_{f_4} g_{A_1 f_4 \bar f_2}^{R} g_{A_2 f_1 \bar f_4}^{L} - \sum_{f_4} m_{f_1} g_{A_1 f_4 \bar f_2}^{R} g_{A_2 f_1 \bar f_4}^{R} \sin^2\frac{\theta}{2} - \sum_{f_4} m_{f_2} g_{A_1 f_4 \bar f_2}^{L} g_{A_2 f_1 \bar f_4}^{L} \sin^2\frac{\theta}{2} + \frac{1}{2} \sum_{\varphi} v_\varphi g_{A_1 A_2 \varphi} y_{\varphi f_1 \bar f_2}^{LR} \nonumber \\
&& + \frac{1}{2} \sum_{A_3} g_{A_1A_2A_3} \left[ g_{A_3 f_1\bar f_2}^{R} m_{f_1} \left( \cos\theta - \frac{M_{A_1}^2 - M_{A_2}^2}{M_{A_3}^2} \right) + g_{A_3 f_1\bar f_2}^{L} m_{f_2} \left( \cos\theta + \frac{M_{A_1}^2 - M_{A_2}^2}{M_{A_3}^2} \right) \right] = 0 \ .
\end{eqnarray}
\end{footnotesize}
 \noindent There is a similar one as Eq. (\ref{sumrule2}) with $L\leftrightarrow R$ arising from $\mathcal{A}^{(++)}_{f_1\bar f_2\to A_1 A_2}$. One can also rewrite Eq. (\ref{sumrule2}) in a simpler form by using Eq.~(\ref{sumrule1}),
\begin{footnotesize}
\begin{eqnarray}\label{sumrule2a}
\hspace{-1cm}&& \sum_{f_3} m_{f_3} g_{A_1 f_1 \bar f_3}^{L} g_{A_2 f_3 \bar f_2}^{R} + \sum_{f_4} m_{f_4} g_{A_1 f_4 \bar f_2}^{R} g_{A_2 f_1 \bar f_4}^{L} - \sum_{f_4} m_{f_1} g_{A_1 f_4 \bar f_2}^{R} g_{A_2 f_1 \bar f_4}^{R} - \sum_{f_4} m_{f_2} g_{A_1 f_4 \bar f_2}^{L} g_{A_2 f_1 \bar f_4}^{L}  \\
&&=  - \frac{1}{2} \sum_{\varphi} v_\varphi g_{A_1 A_2 \varphi} y_{\varphi f_1 \bar f_2}^{LR} + \frac{1}{2} \sum_{A_3} g_{A_1A_2A_3} \left[ g_{A_3 f_1\bar f_2}^{R} m_{f_1} \left( 1+ \frac{M_{A_1}^2 - M_{A_2}^2}{M_{A_3}^2} \right) + g_{A_3 f_1\bar f_2}^{L} m_{f_2} \left( 1- \frac{M_{A_1}^2 - M_{A_2}^2}{M_{A_3}^2} \right) \right] ~. \nonumber
\end{eqnarray}
\end{footnotesize}

If the sum rule Eq.~(\ref{sumrule2}) is not satisfied, $\mathcal{A}_{\frac{1}{2}}\neq0$, relevant new states must appear below the scale
\begin{eqnarray}\label{sLR}
\Lambda = {\rm Min} \left\{ \frac{32\pi} {\int_{-1}^{1} d (\cos\theta) d_{\lambda 0}^J(\theta) A_{\frac{1}{2}, \lambda}}, \ \ {\rm for\ all\ }J\right\} \ .
\end{eqnarray}
When the rum rule Eq.~(\ref{sumrule2}) is satisfied, $A_{\frac{1}{2}}=0$. All the remaining terms go to zero at infinite energy.
Therefore, when discussing {\it finite amplitudes} at high energy below, we will only talk about the chirality preserving cases, which has the general asymptotic form given Eq.~(\ref{finite}).


For practical purposes, we also present the sum rules Eqs.~(\ref{sumrule1}) and (\ref{sumrule2}) in the limit of massless initial state fermions $f_1$, $f_2$. In this case, one does not have to distinguish between helicity and chirality eigenstates and the sum rules  simplify to
\begin{eqnarray}\label{sumrule1b}
&&\sum_{f_3} g_{A_1 f_1 \bar f_3}^{L} g_{A_2 f_3 \bar f_2}^{L} - \sum_{f_4} g_{A_1 f_4 \bar f_2}^{L} g_{A_2 f_1 \bar f_4}^{L}  = \sum_{A_3} g^L_{A_3 f_1 \bar f_2} g_{A_1A_2A_3}  \ ,
\end{eqnarray}
and
\begin{eqnarray}\label{sumrule2b}
&&\sum_{f_3} m_{f_3} g_{A_1 f_1 \bar f_3}^{L} g_{A_2 f_3 \bar f_2}^{R} + \sum_{f_4} m_{f_4} g_{A_1 f_4 \bar f_2}^{R} g_{A_2 f_1 \bar f_4}^{L} + \frac{1}{2} \sum_{\varphi} v_\varphi g_{A_1 A_2 \varphi} y_{\varphi f_1 \bar f_2}^{LR} = 0 \ .
\end{eqnarray}
Meanwhile, the finite amplitude Eq.~(\ref{finite}) also simplifies to
\begin{footnotesize}
\begin{eqnarray}\label{finite2}
\mathcal{A}_0 &=& - \sum_{f_3} g_{A_1 f_1 \bar f_3}^{L} g_{A_2 f_3 \bar f_2}^{L}\frac{M_{A_1}^2 + M_{A_2}^2 - m_{f_3}^2 \csc^2 \frac{\theta}{2} }{2M_{A_1} M_{A_2}} \sin\theta + \sum_{f_4} g_{A_1 f_4 \bar f_2}^{L} g_{A_2 f_1 \bar f_4}^{L} \frac{M_{A_1}^2 + M_{A_2}^2 - m_{f_4}^2 \sec^2 \frac{\theta}{2}}{2M_{A_1} M_{A_2}} \sin\theta \nonumber \\
&&\ + \sum_{A_3} g^L_{A_3 f_1 \bar f_2} g_{A_1A_2A_3} \frac{  M_{A_3}^2 }{2M_{A_1} M_{A_2}} \sin\theta, \ \ \ (\theta\neq 0, \pi) \ .
\end{eqnarray}
\end{footnotesize}
Before closing this section, we illustrate how the above sum rules work in two explicit examples where unitarity has been well established.

\subsection{Application to standard model}

The first example is the SM. We consider amplitudes for massless quark-antiquark scattering into gauge bosons. The final gauge bosons include $\gamma$, $Z$ and $W^\pm$.
The scattering of $q \overline{q} \to \gamma\gamma, \gamma Z, ZZ$ occurs through $q$ exchange in both t-, u-channels, and without chirality flip. Since all coupling in t- and u-channels are identical, the sum rule Eq.~(\ref{sumrule1b}) is satisfied.
The scattering $u_L\overline{d}_L \to ZW^+, \gamma W^+$ can occur through t-channel $d_L$ exchange, u-channel $u_L$ exchange and s-channel $W^{+*}$ exchange. The sum rule is realized since
$\frac{g}{\sqrt2} \frac{g}{\cos\theta_W} \left[(T_{3L} - Q \sin^2\theta_W)_u - (T_{3L} - Q \sin^2\theta_W)_d \right] = \frac{g}{\sqrt2} g \cos\theta_W$. Finally, the scattering $q\overline{q} \to W^+W^-$ occurs through t-channel $q$ exchange and s-channel $\gamma, Z$ exchange. The sum rule is realized as
$g^2 (T_{3L})_q =  \frac{g}{\cos\theta_W} (T_{3L} - Q \sin^2\theta_W)_q\ g\cos\theta + e^2 Q_q$.

\subsection{Application to seesaw models of neutrino mass}

The second application we discuss is the seesaw models for neutrino mass.
We will consider the lepton number violating process $e^-e^- \to W^-W^-$, which ought to be proportional to the Majorana mass~\cite{Maltoni:2000iq}. Thus it belongs to the class of scatterings with chirality flip, i.e., it has the general form of Eq.~(\ref{massinsertion}). If neutrino Majorana mass is the only source of lepton number violation, the process can occur through t-channel or u-channel neutrino exchange, with a Majorana mass insertion. However, as shown in the sum rule Eq.~(\ref{sumrule2}), in this case the t- and u-channel amplitudes have the same sign, thus no cancellation occurs among the two. In this case, the asymptotic form of the scattering amplitude would increase with energy
$\mathcal{A}^{(\nu)}({e^-e^- \to W^-W^-}) \simeq {2g^2 m_\nu} \sqrt{s}/{M_{W}^2}$. Using Eq.~(\ref{sLR}), unitarity tells us high scale physics for neutrino mass generation must appear below the scale
$\Lambda \lesssim {16\pi M_{W}^2}/({g^2 m_\nu}) \simeq 10^{16}\,{\rm GeV}$~\cite{Maltoni:2000iq}.
In fact, gauge invariant Majorana neutrino masses in the minimal SM must arise from the Weinberg operator $(LH)^2/\Lambda$. With this operator, there is also an s-channel SM Higgs boson exchange contribution to the scattering, which restores unitarity.

Next, we show how unitarity is restored in three types of seesaw theories. In type I and type III seesaw models, neutrino masses arise from the mixing of the light neutrino with a heavy Majorana fermion $N$. The mixing angle is $\xi = \sqrt{|m_\nu/m_N|}$. Because of this mixing, when scattering energy is higher than these heavy fermion mass, it also contributes to the above lepton number violating process. The corresponding amplitude is
$\mathcal{A}^{(N)}({e^-e^- \to W^-W^-}) \simeq {2g^2\xi^2 m_N} \sqrt{s}/{M_{W}^2}$.
The type I seesaw mass formula $m_\nu = - \xi^2 m_N$ contains a minus sign which facilitates the cancellation with t- and u-channel contributions and restores unitarity.
On the other hand, in the case of type II seesaw, neutrino mass arises from coupling of the light neutrino to the VEV of a complex Higgs triplet, $m_\nu = y_\Delta v_\Delta$. Due to gauge invariance, the above scattering can also happen through s-channel exchange of doubly-charged component from the Higgs triplet, $\mathcal{A}^{(\Delta)}({e^-e^- \to W^-W^-}) \simeq - {2y_\Delta g^2 v_\Delta} \sqrt{s}/{M_{W}^2}$, which also exactly cancels the t- and u-channel contributions.

Now that we have verified the validity of the general sum rules in two known cases, we turn to theoretical constraints  for standard model extensions with additional $W'^\pm$ gauge boson.

\section{Unitarity constraints on general $\bg{W'}$ models}\label{genaralW'}

In this section we consider simple extensions of the standard model with an additional $W'$ gauge boson which couples to SM fermions.
We focus on the scattering amplitudes for fermions to gauge bosons. In particular, we point out that generically a $Z'$ gauge boson also has to be present in order to preserve perturbative unitarity of all the amplitudes. The coupling of $Z'$ to the fermions are fixed by the sum rules with its mass bounded from above, as we shall see.

We start by considering the scattering $f_1\bar f_1\to W'^+W'^-$, where $W'$ is assumed to possess the interaction $g' \bar f_1 f_3 W'^+ +{\rm h.c.}$ with  $f_1$, $f_3$ being standard model fermions with electric charges $Q_{f_1} = Q_{f_3}+1$. With this coupling, such processes can take place through t-channel $f_2$ exchange.  This amplitude itself would violate unitarity at high energy.
One possible solution, according to the sum rule, is to introduce corresponding u-channel process. In order to preserve electric charge, the new fermion $f_4$ to be exchanged in the u-channel process must have electric charge $Q_{f_4}=Q_{f_1}+1$ and an identical coupling strength in $g' \bar f_4 f_1 W'^+ +{\rm h.c.}$ However, in this case one can also consider the scattering $f_4\bar f_4\to W'^+W'^-$ and new fermion with higher electromagnetic charge must also be introduced. Therefore, unless we introduce infinite chain of fermions whose electric charge differ by one unit, unitarity cannot be restored.\footnote[1]{The above argument is based on the assumption where the fermions $f_1$, $f_2$ and $f_3$ are all SM fermions, which is the case in the conventional $W'$ models. We will show that the infinite tower of particles can be avoided in generalized models in Section~\ref{exception} .}

Therefore, in order to find a finite solution, one must rely on s-channel processes. In fact, there could be $\gamma W'W'$ and $ZW'W'$ couplings which also contribute to the above process. We shall see that such couplings are fixed by considering various gauge boson final states and additional s-channel $Z'$ contribution is generally necessary.

In the following, we also take into account possible deviations from theoretical predictions of SM couplings, due to possible $W-W'$ and $Z-Z'$ mixings. We do not include possible small deviations from the SM in $Wf\bar f'$ coupling due to the precise measurement of $G_F$ and $M_W$, or deviations in $\gamma f\bar f$, $\gamma WW$ couplings which have their forms guaranteed by
electromagnetic gauge invariance. We parametrize the $ZWW$ and $Zf\bar f$ couplings as follows.
\begin{eqnarray}
&&g_{ZWW} = (1+\delta_1) g \cos\theta_W, \nonumber \\
&&g_{Zf\bar f} = \frac{g}{\cos\theta_W} \left[(1+\delta_2) T_{3L} - (1+\delta_3)Q \sin^2\theta_W + \delta_{4f} \right] \,.
\end{eqnarray}
Here $\delta_{4f}$ parametrizes the deviation of $Z$ coupling to RH fermions, due to $Z-Z'$ mixing.
We will present our results to leading order in the small parameters $\delta_i$, which are allowed to be at most 5\%~\cite{:2003ih}.

\subsection{General left-handed $\bg{W'}$}

We first consider the case where $W'$ couples to left-handed SM quark doublets, in a way similar to the $W$-boson:
\begin{eqnarray}
\mathcal{L} = \frac{g'}{\sqrt2} \bar u \gamma^\mu P_{L} d W'^+_{\mu} + {\rm h.c.} \ .
\end{eqnarray}
We will also include a $Z'$ gauge boson and consider all possible couplings among gauge bosons as well as general couplings of the $Z'$ to SM fermions. We show the relevance of $Z'$ due to its non-vanishing couplings demanded by the sum rules. Throughout the paper, we do not consider the coupling of $W'$ to leptons and assume the gauge anomalies are canceled by heavy fermion spectators~\cite{Preskill:1990fr}, when embedded in a complete theory.

The processes we consider are summarized in Table~\ref{table1}.
\begin{table}[b!]
\centerline{\begin{tabular}{|l|l|l|}
\hline
Process       &  Exchanged particle &  Sum rule \\
\hline\hline
& t-channel $d$ quark  &   \\
\raisebox{2.0ex}[0pt]{$u_L \bar u_L\to W^+W^-$} & s-channel $\gamma$, $Z$, $Z'$ &\raisebox{2.0ex}[0pt]{$g_{Z'WW} = \mathcal{O}(\delta_i)$
} \\
\hline
 & t-channel $d$ quark  &  \\
$u_L\bar d_L\to W^+ Z$ & u-channel $u$ quark & $g_{ZWW'} = \mathcal{O}(\delta_i)$ \\
& s-channel $W$, $W'$ & \\
\hline
 & t-channel $d$ quark & \\
$u_L\bar d_L\to W'^+ Z$ & u-channel $u$ quark & $g_{ZW'W'} = g \cos\theta_W + \mathcal{O}(\delta_i)$ \\
& s-channel $W$, $W'$ & \\
\hline
 &  t-channel $d$ quark &  \\
$u_L\bar d_L\to W^+ Z'$ & u-channel $u$ quark & $g'\ g_{Z'WW'} = g\ (g_{Z' u_L\bar u_L} - g_{Z' d_L\bar d_L})$ \\
& s-channel $W$, $W'$ & \\
\hline
 & t-channel $d$ quark  &  \\
$u_L\bar d_L\to W'^+ Z'$& u-channel $u$ quark & \raisebox{2.0ex}[0pt]{$g\ g_{Z'WW'} + g'\ g_{Z'W'W'}$}\\
& s-channel $W$, $W'$ & \raisebox{2.0ex}[0pt]{\hspace{1.6cm}$= g'(g_{Z' u_L\bar u_L} - g_{Z' d_L\bar d_L}) + \mathcal{O}(\delta_i)$}\\
\hline
 & t-channel quark if LH $q$ & \raisebox{0.0ex}[0pt]{$\frac{g}{\cos\theta_W} (T_{3L} - Q \sin^2\theta_W) \ g_{ZW'W'}$} \\
\raisebox{2.0ex}[0pt]{$q\overline{q}\to W'^+ W'^-$} & s-channel $\gamma$, $Z$, $Z'$ & \hspace{0.1cm} $+ g_{Z' u_L\bar u_L}\ g_{Z'W'W'} + e^2 Q = g'^2 T_{3L}+ \mathcal{O}(\delta_i)$ \\
\hline
\end{tabular}}
\caption{Processes considered and sum rules, in the case of $W'$ with couplings to left-handed fermions. Here $\bar q_{L} \equiv \overline{(q_{L})}$, etc. Solutions to the sum rules are given in Eq.~(\ref{WLresult}).}
\label{table1}
\end{table}
{\it A crucial point to note is, there are enough sum rules to solve for all the unknown couplings of $Z'$.} We summarize them here, taking into account
the leading order $\delta_i$ corrections.
\begin{eqnarray}\label{WLresult}
g_{Z'WW}  &=& \frac{g^2}{g'} (- \delta_1 - \delta_2) + \mathcal{O}(\delta_i^2), \quad g_{ZWW'} = \frac{g^2}{g'} (\delta_2 \sec\theta_W - \delta_1 \cos\theta_W) + \mathcal{O}(\delta_i^2),
\nonumber \\
g_{ZW'W'} &=& g \cos\theta_W + \mathcal{O}(\delta_i), \quad g_{Z'WW'} = g + \mathcal{O}(\delta_i), \ \ g_{Z'W'W'} = \frac{g'^2 - g^2}{g'} + \mathcal{O}(\delta_i) \nonumber \\
g_{Z'q\bar q} &=& g' T_{3L} + \mathcal{O}(\delta_i),
\end{eqnarray}
along with $\delta_3\approx -\delta_1$ and $\delta_{4q} \approx 0$, in order to balance the coefficients of $T_{3L}$, $Q$ and $T'_q$ in the sum rules. These structures are the common predictions of left-handed $W'$ models.
Recall that $\delta_i\lesssim 5\%$, thus $g_{ZWW'}$ and $g_{Z'WW}$ are only tiny couplings, arising from $W$$-$$W'$ and/or $Z$$-$$Z'$ mixings.

Since the $Z'$ must possess non-zero couplings to fermions and other gauge bosons, it cannot be simply decoupled from a consistent theory. We would like to emphasize that when the $W'$-fermion coupling $g'$ is given, the $Z'$-fermion coupling is fixed automatically. The $W'_L$ model studied in Ref. \cite{Schmaltz:2010xr} is one of the incarnations of this generic structure. In particular, the model starts from $SU(2)_1\times SU(2)_2\times U(1)_Y$ where the two $SU(2)$'s break down to the diagonal subgroup which is identified as the SM $SU(2)_L$. The SM fermion doublets are charged only under the first $SU(2)$.
After this stage of symmetry breaking, the $W'$, $Z'$ couplings to fermions  has a strength given by
$g_1^2 T_{3L}/\sqrt{g_1^2+g_2^2}$, where $g_{1,2}$ are gauge couplings corresponding to $SU(2)_1$ and $SU(2)_2$, respectively.

Next, we proceed to calculate the finite amplitude at very high energy, given the above sum rules. We are interested in finding the upper bound on the $Z'$ mass scale required by perturbative unitarity. Among the above processes, the most important one is $q\overline{q}\to W'^+ W'^-$. Using the general formula Eq.~(\ref{finite}), we get
\begin{eqnarray}
\mathcal{A}_0 &=& \frac{1}{2}\left[ g'^2 T_{3L} \left( \frac{M_{Z'}^2}{M_{W'}^2} - 2 \right) - g^2 T_{3L} \frac{M_{Z'}^2}{M_{W'}^2}  + g^2 \frac{M_{Z}^2}{M_{W'}^2} \left( T_{3L} - Q \sin^2\theta_W \right) \right]\sin\theta \ .
\end{eqnarray}
Here we neglect all the quark masses. In this case, the amplitude is a pure p-wave.
The partial wave amplitude can be calculated with Eq.~(\ref{partial}).

As an example, we consider the initial quarks being  $u_L$ and $\bar u_L$ and the fermion being exchanged in the t-channel being the $d$ quark.
Demanding the partial wave $|a_{\lambda=1}^{J=1}|<1$, the allowed regions in the $g'-M_{Z'}$ parameter space for different $W'$ masses equal to 150, 500 and 800\,GeV are shown in the left panel of Fig.~\ref{WL}.

The first point we note from the plot is the upper bound on $Z'$ mass is most stringent for very large coupling $g'$.
This will be important in obtaining the $Z'$ mass bound in top quark asymmetry models  studied in the next section.

There is also a possibility $g'=g$ where one can have the $Z'$ much heavier than the $W'$. This can be understood because in this case $W'$ couples exactly
in the same way as the SM $W$ boson, and the SM itself is a consistent theory from the unitarity point of view.
Perturbative unitarity of the finite part of $u_L\bar d_L\to W'^+Z'$ and $u_L\bar d_L\to W^+Z'$ still gives a finite but very mild upper bound
\begin{eqnarray}
M_{Z'} < {\rm Min} \left\{ \frac{96 \pi M_{W'}}{g'^2},~\frac{192 \pi M_{W}}{g'^2} \right\}.
\end{eqnarray}


\subsection{Comment on CDF $\bg{Wjj}$ event excess}

It has been argued that the CDF $Wjj$ excess can be explained via the associated production of $W$ and extra gauge bosons, such as $pp\to W^\pm W'^\mp\to W^\pm jj$, with $M_{W'}\approx 150\,$GeV and a relatively small $g'<g$. The $W'$ needs to be left-handed in order to avoid chiral suppression and to give large enough cross section $\sim$\,4\,pb$^{-1}$. In the $W'_L$ model, unitarity implies that there is a destructive interference between t-channel quark exchange and s-channel $Z'$ exchange. Here we point out that for the model presented in Ref. \cite{Wang:2011uq} to be complete, there must exist some other resonances at low energy, or the $Z'$ coupling should be modified.

\begin{figure}[t!]
\begin{center}
\includegraphics[width=7cm]{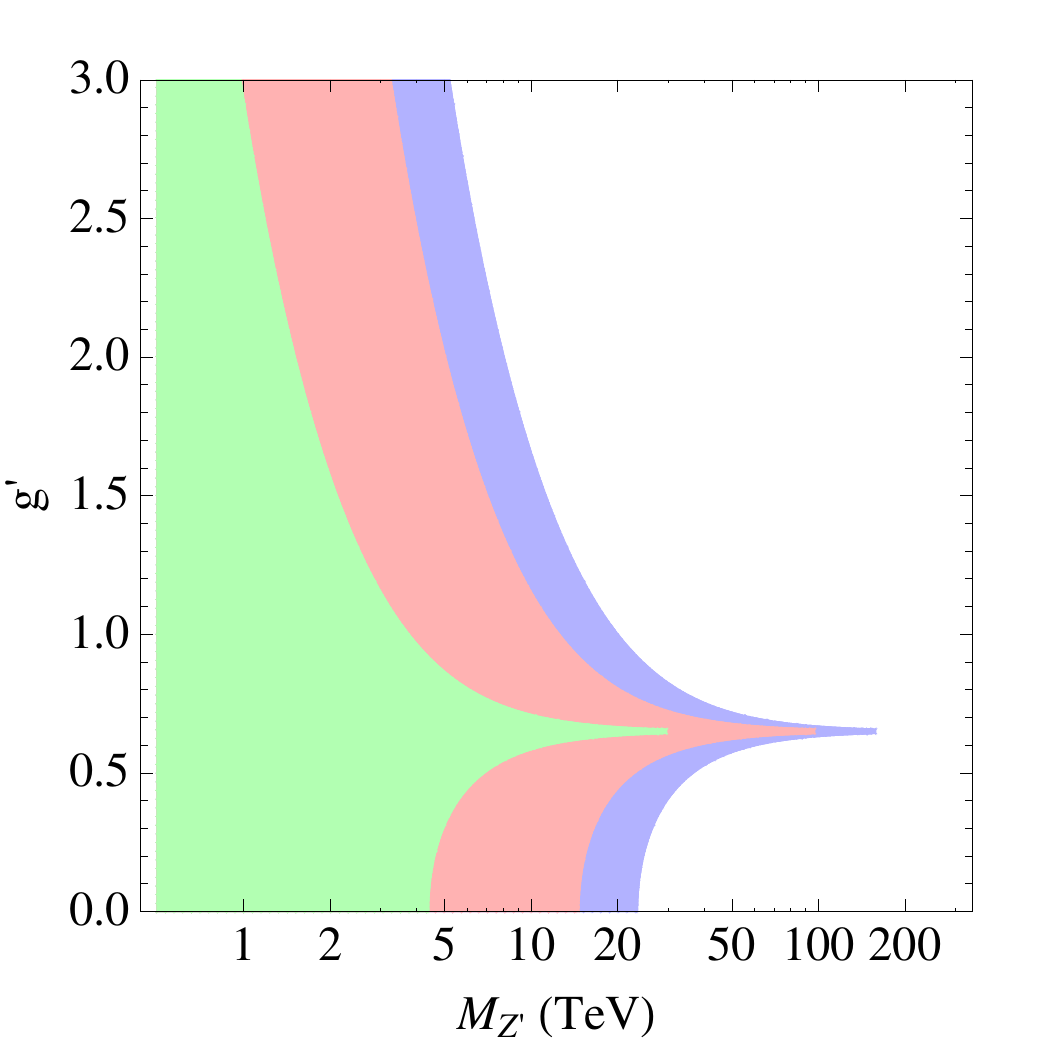} \hspace{0.cm}
\includegraphics[width=7cm]{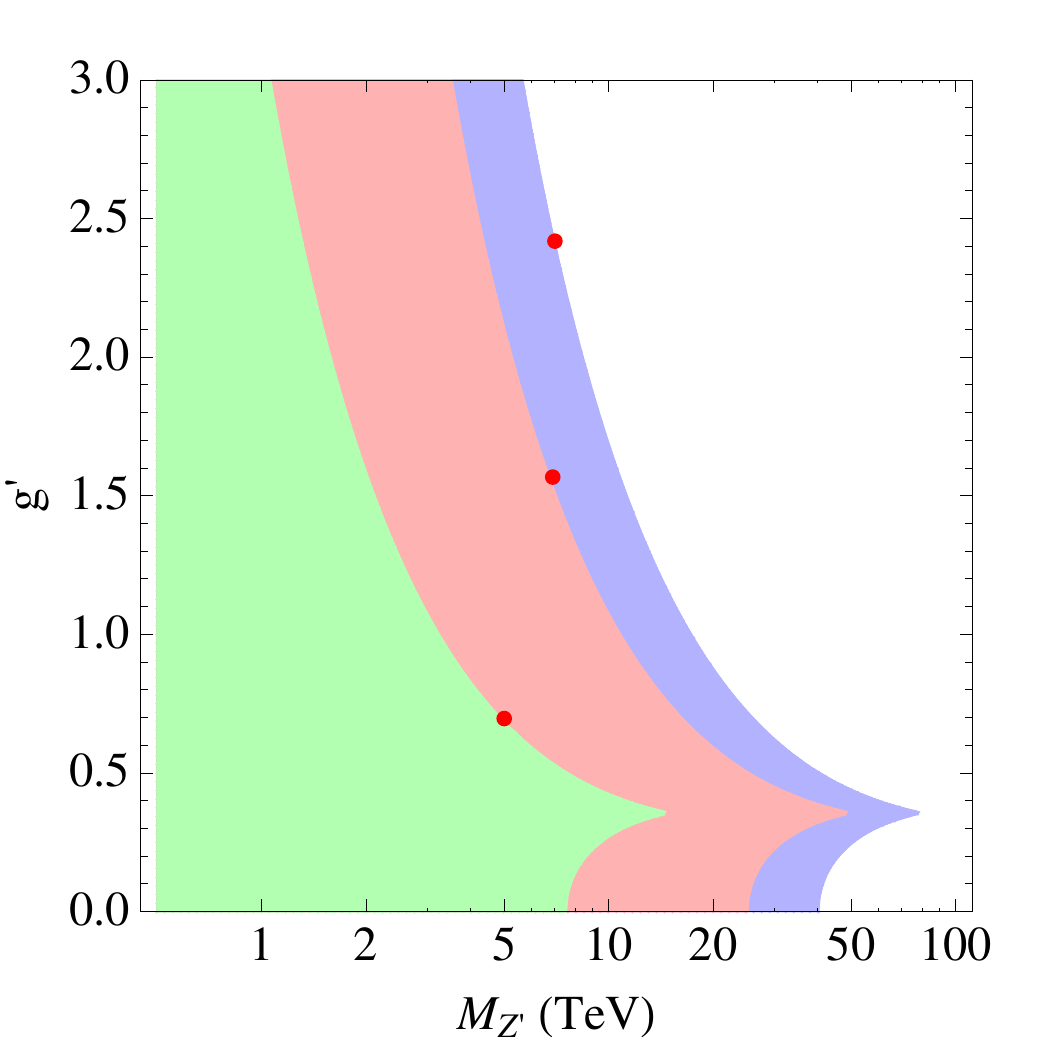}
\caption{Colored regions are allowed in the $g'-M_{Z'}$ parameter space for $M_{W'}=150\,$GeV (green), 500\,GeV (red) and 800\,GeV (blue). Left panel: left-handed $W'$ case; Right panel: right-handed $W'$ case. In the right panel, the red points represent the central values of $(g', M_{W'})$ for explaining the top quark asymmetry where $W'$ couples to $t, d$, to be discussed in the next section. In such models, the upper bound on $Z'$ mass is only a few TeV.}\label{WL}
\end{center}
\end{figure}

From Fig.~\ref{WL}, we learn that the appearance of $Z'$ can be postponed up to $4-5$\,TeV for $g'<0.5$. If $Z'$ mass is close to this upper bound, one can achieve a large hierarchy between the cross sections $\sigma(pp\to W^\pm W'^\mp)$ and $\sigma(pp\to Z W'^\pm)$.
The latter process does not show any excess as reported by the CDF collaboration. It happens through t- and u-channel quark exchange as well as s-channel $W'$ exchange. Since here the $W'$ mass is fixed, the destructive interference for $ZW'$ associated production is much more significant, which leads to a much smaller cross section. In fact, we find for the parameters $M_{W'}=150\,$GeV, $g'=0.3$, and a heavy $Z'$ inaccessible at the Tevatron energy, the leading-order cross sections  calculated using {\sf MadGraph}~\cite{Alwall:2011uj} to be
\begin{eqnarray}
\sigma(pp\to W^\pm W'^\mp) \approx 3.7\,{\rm pb}, \ \ \sigma(pp\to Z W'^\pm) \approx 0.14\,{\rm pb} \ ,
\end{eqnarray}
while the two cross sections are similar for $M_{Z'}\sim M_{W'}$.
Alternatively, since the CDF analysis does not veto the possibility of $Wjj$ being reconstructed into a resonance, one can obtain a large ratio
$\sigma(pp\to W^\pm W'^\mp)/\sigma(pp\to Z W'^\pm)$
by taking a relatively light $Z'$ with appropriate couplings, which can be produced on shell and decays into $W$ and $W'$.

\subsection{General right-handed $\bg{W'}$}

Following similar procedure as in the case of left-handed $W'$, we now study $W'$ with the right-handed couplings:
\begin{eqnarray}
\mathcal{L} = \frac{g'}{\sqrt2} \bar u \gamma^\mu P_{R} d W'^+_{\mu} + {\rm h.c.} \ .
\end{eqnarray}
We consider various processes and obtain the sum rules, as summarized in Table~\ref{table2}.
\begin{table}[h!]
\centerline{\begin{tabular}{|l|l|l|}
\hline
Process       &  Exchange particle &  Sum rules \\
\hline \hline
 & t-channel $d$ quark  &   \\
\raisebox{2.0ex}[0pt]{$u_L\bar u_L\to W^+W^-$} & s-channel $\gamma$, $Z$, $Z'$ &\raisebox{2.0ex}[0pt]{$g_{Z'WW} =  \mathcal{O}(\delta_i)$} \\
\hline
$u_R\bar d_R\to W^+ Z$ & s-channel $W'$ & $g_{ZWW'} = 0$ \\
\hline
 & t-channel $d$ quark & \\
$u_R\bar d_R\to W'^+ Z$ & u-channel $u$ quark & $g_{ZW'W'} = - g \frac{\sin^2\theta_W}{\cos\theta_W} + \mathcal{O}(\delta_i)$  \\
& s-channel $W'$ & \\
\hline
&  t-channel $d$ quark & \\
$u_L\bar d_L\to W^+ Z'$ & u-channel $u$ quark &  $g_{Z' u_L\bar u_L} = g_{Z' d_L\bar d_L} + \mathcal{O}(\delta_i)$  \\
 & s-channel $W$ & \\
\hline
$u_R\bar d_R\to W^+ Z'$ & s-channel $W'$ & $g_{Z'WW'} =  0$ \\
\hline
 & t-channel $d$ quark  &  \\
$u_R\bar d_R\to W'^+ Z'$& u-channel $u$ quark & $g'\ (g_{Z' u_R\bar u_R} - g_{Z' d_R\bar d_R}) =g'\ g_{Z'W'W'} + \mathcal{O}(\delta_i)$\\
& s-channel $W'$ & \\
\hline
 & t-channel quark if RH $q$ & $\frac{g}{\cos\theta_W} (T_{3L} - Q \sin^2\theta_W)_q \ g_{ZW'W'}$ \\
\raisebox{2.0ex}[0pt]{$q\overline{q}\to W'^+ W'^-$} & s-channel $\gamma$, $Z$, $Z'$ &\raisebox{0.0ex}[0pt]{\hspace{0.8cm}$g_{Z' q\bar q}\ g_{Z'W'W'}  +  e^2 Q_q = g'^2 T'_{q} + \mathcal{O}(\delta_i)$} \\
\hline
\end{tabular}}
\caption{Processes considered and sum rules, in the right-handed $W'$ case. Solutions to this set of sum rules are given in Eq.~(\ref{WRcouple}).}
\label{table2}
\end{table}

Again, all the relevant couplings can be solved with the listed sum rules.
\begin{eqnarray}\label{WRcouple}
g_{Z'WW} &=& \frac{\sqrt{g'^2-g^2\tan^2\theta_W}}{\tan^2\theta_W} (-\delta_1 - \delta_2) + \mathcal{O}(\delta_i^2), \ \ g_{ZWW'} = g_{Z'WW'} = 0,  \nonumber \\
g_{ZW'W'} &=& - g \frac{\sin^2\theta_W}{\cos\theta_W} + \mathcal{O}(\delta_i),
\quad
g_{Z'W'W'} = \sqrt{g'^2 - g^2\tan^2\theta_W} + \mathcal{O}(\delta_i), \nonumber \\
g_{Z'q\bar q} &=& \frac{g'^2 T'_{q} + g^2\tan^2\theta_W (T_{3L}-Q)}{\sqrt{g'^2 - g^2\tan^2\theta_W}}  + \mathcal{O}(\delta_i) \ .
\end{eqnarray}
Here, $T'_{q_L}=0$ and $T'_{u_R}=-T'_{d_R}=1/2$. The left-right symmetric model is a well-known example for this scenario, where right-handed quarks form doublets and $T'_q$ is identified as $T_{3R}$. The $g_{Z'q\bar q}$ coupling derived here from unitarity agrees with that given in Ref. \cite{Nemevsek:2011hz}.

The sum rules also imply the following relations among the small parameters, in the right-handed $W'$ case:
$\delta_2 - \delta_3 \sin^2\theta_W + \delta_1 \cos^2\theta_W = 0$ and $\delta_4 g^2 \tan^2\theta_W = (\delta_1 + \delta_2) g'^2$, where we defined $\delta_{4q}\equiv\delta_4 T'_{q}$.

We point out an interesting difference between right-handed and left-handed $W'$ models. The $g_{Z'WW'}$ coupling is vanishing in the former case but is nonzero in the latter case.
There is thus a possibility to distinguish the two class of models by studying $Z'\to W^\pm W'^\mp$ process.

Next, we calculate the finite part of $q\overline{q}\to W'^+ W'^-$ amplitude at very high energy, when the above sum rules are satisfied. Using the general formula Eq.~(\ref{finite}), we get
\begin{small}
\begin{eqnarray}\label{AR}
\mathcal{A}_0 &=& \frac{1}{2} g'^2 T'_{q} \left( \frac{M_{Z'}^2}{M_{W'}^2} - 2 \right) \sin\theta  \nonumber \\
&& + \left[g^2 \tan^2 \theta_W (T_{3L} - Q ) \frac{M_{Z'}^2}{2 M_{W'}^2} - g^2 \tan^2 \theta_W (T_{3L} - Q \sin^2\theta_W) \frac{M_Z^2}{2 M_{W'}^2} \right] \sin\theta.
\end{eqnarray}
\end{small}
Again we neglect the masses of quarks being exchanged in the t-channel, in which case the amplitude is pure p-wave. Demanding the partial wave $|a_{\lambda=1}^{J=1}|<1$, we find the allowed region in $g'-M_{Z'}$ parameter space for different $W'$ masses equal to 150, 500 and 800\,GeV as shown in the right panel of Fig.~\ref{WL}. We have considered unitarity of both $u_R\overline{(u_R)}\to W'^+ W'^-$ and $d_R\overline{(d_R)}\to W'^+ W'^-$.

\subsection{Fermion mass limits from unitarity}

In many extensions of the SM, the presence of a $W'$ and a $Z'$ are often accompanied by exotic fermions.  These exotic fermions typically interact with SM fermion though the $W'$. A well known example is the left-right symmetric model in which $W'$ couples to the right-handed neutrino $N_R$ and the right-handed electron.  In the alternate left-right model \cite{Babu:1987kp}, which has the same gauge structure as the left--right model but has an extended fermion
sector from $E_6$ embedding, the $W'$ couples to the right-handed up-quark and an exotic down-quark $d'_R$. The unitarity argument can be
used to derive upper limits on the masses of these exotic fermions $N_R$ and $d'_R$ as we now discuss.  While there
are other processes, such as $f \bar{f} \rightarrow f \bar{f}$ mediated by gauge bosons, which may provide better limits on the fermion masses,
here we use the sum rules of Eq. (\ref{finite}) with $L \leftrightarrow R$ interchange to derive limits, which are nontrivial.

Let us consider $f\bar{f} \to W'^+W'^-$ process with $f$ being the right-handed electron (right-handed up-quark) in the LR model (alternate LR model). Since the internal t-channel fermion is heavy, the finite part of the amplitude given in Eq. (\ref{AR}) should be altered to
\begin{eqnarray}
\mathcal{A}_0(f\bar{f} \to W'^+W'^-) &=& \left[ \frac{1}{2}g'^2 T'_{f}\left(\frac{M_{Z'}^2}{M_{W'}^2}-2\right)  + g^2 \left(Q_f  \tan^2\theta_W\sin^2\theta_W  \frac{M_Z^2}{2M_{W'}^2} - Q_f \tan^2\theta_W \frac{M_{Z'}^2}{2M_{W'}^2} \right) \right.  \nonumber \\
&& \left. + g'^2T'_{f} \frac{m_{f'}^2 \csc^2\tfrac{\theta}{2}}{2M_{W'}^2}\right] \sin\theta,
\label{modified}
\end{eqnarray}
where $m_{f'}$ is the internal fermion mass. For our numerical estimate we take $g'=g$, which is justified if parity is a good symmetry. In the minimal version of the models being considered (LR or alternative LR), there exist $W'-Z'$ mass relations given by
\begin{eqnarray}
\frac{M_{Z'}}{M_{W'}} = a, \quad {\rm with} \quad a = \left\{ \begin{array}{ll} \frac{\sqrt{2}\cos\theta_W}{\sqrt{\cos 2\theta_W}}
 & {\rm in~LR~model} \\ \\
\frac{\cos\theta_W}{\sqrt{\cos 2\theta_W}} & {\rm in~alternate~LR~model} \end{array} \right.
\end{eqnarray}
Because this mass ratio is fixed, unitarity requirement would give an upper limit on $m_{f'}$ at very large $M_{W'}$. Thus, one can neglect the $M_Z/M_{W'}$ terms in Eq. (\ref{modified}). By requiring the partial wave amplitude $|a_1^1|<1$, we get the limit
\begin{eqnarray}
m_{f'}^2 <  \frac{M_{W'}^2}{3\sqrt{2}g^2} \left| \frac{96\pi}{T'_{f}} + \sqrt{2}g^2
\frac{(2-a^2)T'_{f} + a^2Q_f \tan^2\theta_W}{T'_{f}} \right|.
\end{eqnarray}
By substituting $T'_{e_R}=-1/2,~Q_e=-1$ for $e_R$ and $T'_{u_R}=1/2,~Q_u=2/3$ for $u_R$, we obtain
\begin{eqnarray}
m_{N_R} < 18.33 \,M_{W'} \quad{\rm and} \quad  m_{d'_R} < 18.35\, M_{W'}~.
\end{eqnarray}

\section{Implications of unitarity on $\bg{W'}$ models for top quark forward-backward asymmetry}

The top quark forward-backward asymmetry has been measured by both D0 and CDF collaborations~\cite{:2007qb, Aaltonen:2011kc} and shows more than $3\sigma$ deviation from the SM prediction in the high $t\bar t$ invariant mass frame~\cite{Kamenik:2011wt}. Meanwhile, the measurement of top pair production cross section agrees with SM within $1\,\sigma$. There have been extensive studies of new physics scenarios to explain the asymmetry, amongst which t-channel new gauge boson exchanges seem to be popular\cite{Gresham:2011pa, Gresham:2011fx}. It induces a large forward-backward asymmetry thanks to the Rutherford singularity~\cite{Jung:2009jz} and leads to small modification to the top quark production cross section. While the $Z'$ boson exchange also predicts large same-sign top quark production, in its minimal realization this explanation would appear to be in contradiction with LHC results~\cite{Berger:2011ua, AguilarSaavedra:2011zy, Chatrchyan:2011dk, Ko:2011vd}.  The $W'$ models are free from such constraints and remain as a valid explanation. In this section we will apply the consideration of perturbative unitarity developed here to the $W'$ models for top asymmetry, and study the phenomenological consequences.

\subsection{Left-handed $\bg{W'}$}

We first study the following couplings of $W'$ gauge boson to left-handed quarks.
\begin{eqnarray}\label{ttbarWL}
\mathcal{L} = \frac{g'}{\sqrt2} \bar t \gamma^\mu P_{L} d W'^+_{\mu}  + \frac{g''}{\sqrt2} \bar u \gamma^\mu P_{L} b W'^+_{\mu} + {\rm h.c.} \ .
\end{eqnarray}
If we neglect for simplicity any mixing between SM quarks with heavy exotic quarks, then $g''=g'$ due to $SU(2)_L$ gauge invariance. With the flavor changing coupling in the first term, the top quark pair production $d\bar d\to t\bar t$ receives t-channel $W'$ contribution, which enhances the forward-backward asymmetry. We note that the same coupling also leads to $d\bar d\to W'^+ W'^-$ though t-channel top quark exchange. Hence the perturbative unitarity constraints discussed in the previous sections would apply here.
We will show that in addition to the sum rules obtained in Eq.~(\ref{WLresult}), there are further constraints due to the flavor changing nature of the $W'$ coupling.

We first consider the $u_L\bar t_L\to W^+ W'^-$, $d_L \bar b_L\to W^+ W'^-$ scattering and their charge conjugate processes. With the above coupling, these processes can only happen in t-channel or u-channel, unless there are flavor-changing $Z'$ couplings to quarks (which are highly
constrained). The sum rule for this process can be written as
$g_{Z'WW'} g_{Z' u_L\bar t_L} = - g_{Z'WW'} g_{Z' d_L\bar b_L} = {g g'}/{2}$. The same flavor-changing $Z'$ coupling will also unitarize the behavior of $t_L\bar u_L\to W^- W'^+$ and $b_L \bar d_L\to W^- W'^+$ scatterings, as long as the second term of Eq.~(\ref{ttbarWL}) is present.

Next, we consider the process $u_L\bar d_L\to W'^+ Z'$. In this model, it can occur not only through s-channel $W^+$ exchange, but also
 through t-channel $b_L$ and u-channel $t_L$ exchange. The corresponding sum rule is
$g\ g_{Z'WW'} = g'\ (g_{Z'u_L\bar t_L} - g_{Z'd_L\bar b_L})$. Together with the sum rules obtained in the previous paragraph, we find $g_{Z' u_L\bar t_L} = {g}/{2} $, $g_{Z' d_L\bar b_L} = - {g}/{2} $ and $g_{Z'WW'} = g'$.
On the other hand, one can also consider the process $t_L\bar d_L\to W^+ Z'$ occurring through s-channel $W'^+$ and t-channel $b_L$ and u-channel $u_L$ exchanges, which leads to
$g'\ g_{Z'WW'} = g\ (g_{Z'u_L\bar t_L} - g_{Z'd_L\bar b_L})$, and in turn
$g_{Z' u_L\bar t_L} = {g'}/{2}$, $g_{Z' d_L\bar b_L} = - {g'}/{2}$, $g_{Z'WW'} = g$.
The two solutions for flavor-changing $Zf\bar f'$ couplings can be valid only if $g=g'$, i.e., unitarity demands such left-handed $W'$ must possess the same couplings as the SM $W$.

Combining the above results with Eq.~(\ref{WLresult}), we get
\begin{eqnarray}
&&g_{ZWW'}= g_{Z'W'W'} = g_{Z'WW} = 0, \ \ g_{ZW'W'} = g \cos\theta_W \ , \nonumber \\
&&g_{Z'WW'} = g, \ \ g_{Z'q\bar q} = 0, \ \ g_{Z' u_L\bar t_L} = -  g_{Z' d_L\bar b_L} = \frac{1}{2} g =  \frac{1}{2} g' \ .
\end{eqnarray}
With $W'td$ coupling $g'=g$, we must have a light $W'$ with mass below 200\,GeV in order to fit the top quark asymmetry. Such a mass and coupling seem to be favored by a recent analysis~\cite{Jung:2011id}. Furthermore, one can write down the leading order CKM matrix analog for $W'$ coupling to different flavors of quarks:
\begin{eqnarray}
V'=\left(\begin{array}{ccc}
0 & 0 & 1\\
0 & 1 & 0\\
1 & 0 & 0
\end{array}\right) \ .
\end{eqnarray}

There are severe phenomenological constraints on such models of $W'$ with an accompanying $Z'$. First, since the $Z'$ mediates flavor-changing neutral current at tree level, the coupling $g_{Z' d_L\bar b_L}$ would lead to large $B_d-\overline{B_d}$ mixing amplitude. This would imply that the $Z'$ must be heavier than about 100\,TeV~\cite{D'Ambrosio:2002ex}. As discussed in Fig.~\ref{WL}, such a heavy $Z'$ is barely consistent with the upper bound from unitarity, even for the case of equal couplings $g=g'$.
Second, it has been shown in Ref.~\cite{Craig:2011an} that $W'$ couplings to both $\bar t d$ and $\bar u b$ would lead to much larger single top production cross section at the LHC than what has been observed. We conclude that in its minimal setup, the left-handed $W'$ scenario is less likely to be the explanation for top quark asymmetry.

\subsection{Right-handed $\bg{W'}$}

We next consider $W'$ gauge boson coupling to right-handed quarks:
\begin{eqnarray}
\mathcal{L} = \frac{g'}{\sqrt2} \bar t \gamma^\mu P_{R} d W'^+_{\mu} + {\rm h.c.} \ .
\end{eqnarray}
In this case, we find unitarity does not bring more constraints than those listed in Eq.~(\ref{WRcouple}).
The scattering $u\bar t, d\bar b\to W^+ W'^-$ involves a mass insertion, which according to the sum rule, has to be cancelled by s-channel $\varphi$ exchange (The $Z'$ coupling to quarks is still flavor diagonal and universal). Such a scalar also would lead to tree-level flavor-changing neutral currents and must be made heavy.
This is similar to the problem of the neutral component from the Higgs bidoublet in conventional left-right symmetric models~\cite{Mohapatra:1983ae, Zhang:2007da, Maiezza:2010ic}. Assuming that this constraint is met, right-handed $W'$ can provide a viable explanation to the top quark asymmetry for a wide range of masses and couplings. In order to fit the top quark forward-backward asymmetry, $g'$ has to be larger than the weak coupling $g$, which should increase with increasing $M_{W'}$. Perturbativity of $g'\lesssim \sqrt{4\pi}$ thus implies an upper bound on $M_{W'}$ which is about 1\,TeV.

\begin{figure}[t]
\begin{center}
\includegraphics[width=12cm]{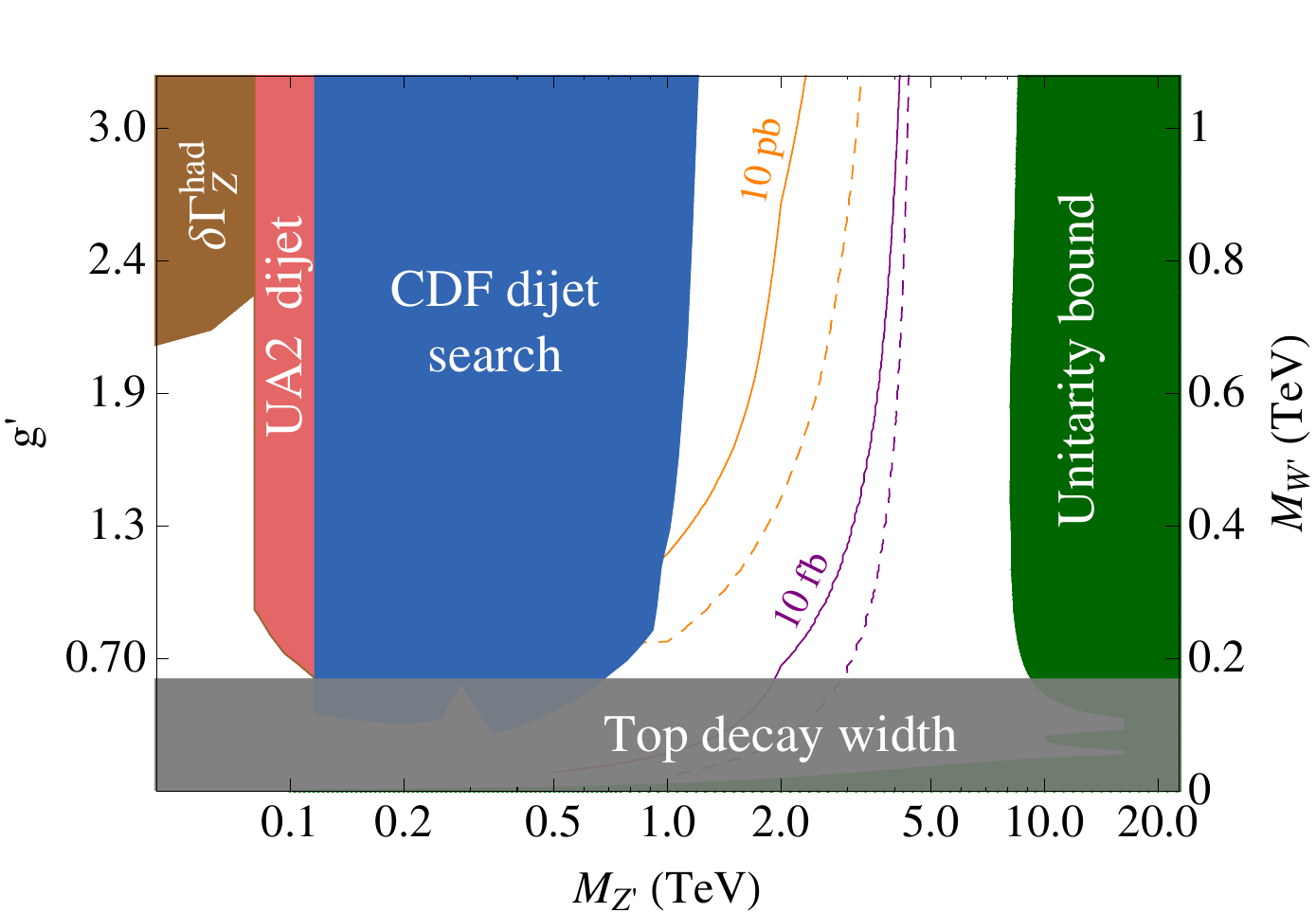}
\caption{Perturbative unitarity and direct search limits on the $Z'$ gauge boson in right-handed $W'$ models for top quark forward-backward asymmetry. We have also shown the contours for $Z'$ production cross sections equal to 10\,fb (purple) and 10\,pb (orange), at 7\,TeV (solid) and 14\,TeV (dashed) LHC energy.
The mass of $W'$ is also shown in accord with $g'$ following the fit given in Ref.~\cite{Shu:2011au}.}\label{WZ}
\end{center}
\end{figure}

We see how unitarity has a strong impact on the $W'$ explanation of top quark asymmetry.
First, unitarity gives an upper bound on the mass of $Z'$, which is
only a few TeV when $g'$ is large, as shown in Fig.~\ref{WL}.
Second,  the $Z'$ coupling strength to fermions is related to that of $W'$, as shown in the right panel of Eq.~(\ref{WRcouple}). A large $g'$ to explain the top asymmetry thus implies a strong $Z'$ coupling to light quarks.
Therefore, most of the allowed parameter space accommodating the $Z'$ could be subject to searches at hadron colliders.
In the model described in Ref. \cite{Jung:2011zv}, the limits from UA2 and Tevatron~\cite{Alitti:1993pn, Aaltonen:2008dn} are shown to have set important constraints. In Fig.~\ref{WZ}, we have included both constraints from unitarity and direct searches. We also demand the $W'$ to be heavier than the top quark in order not to affect the top quark decay rate.

There are basically two regions where right-handed $W'$ models can explain the top quark asymmertry consistently. One region has a very light $Z'$ which has not been well constrained. The other region has the $Z'$ lying in the window between 1 to $\sim$7\,TeV. The Tevatron bound quickly goes away for $M_{Z'}\gtrsim$\,TeV due to the kinematical limit. The LHC is expected to put important constraints on the heavy $Z'$ window~\cite{Salvioni:2009mt}. In Fig.~\ref{WZ}, we also give the contours for production cross sections at 7\,TeV and 14\,TeV LHC. Most of the high $Z'$ mass window can be further probed with high enough luminosity.

\section{Perturbative unitarity in models without a $\bg{Z'}$}\label{exception}

In the class of $W'$ models discussed in previous sections,
a $Z'$ gauge boson of comparable mass was found to be indispensable for restoring perturbative unitarity.
For completeness, in this section we discuss an orthogonal class of $W'$ models where unitarization is realized
without the need for a $Z'$. Here we interpret $W'$ in a more general sense, it will not
be required to have electric charge $\pm 1$.  It does couple with different types of fermions.  If models without a $Z'$ are to contain the SM gauge
symmetry, the extended gauge structure must have rank 4, the same as the rank of the SM gauge group.  Only a finite number of gauge groups
have this property of being rank 4 and containing the SM gauge symmetry.  They are:
(i) $SU(5)$; (ii) $SU(3) \times SU(2) \times SU(2)'$; (iii) $SU(3) \times SU(3)$; (iv) $SU(4) \times SU(2)$; (v) $SO(9)$; (vi) $Sp(8)$; (vii) $F_4$. In fact, models based on these groups which contain the SM symmetry do not have the conventional $W'$ gauge boson with electric harge $\pm1$,
we shall allow for the charge to be arbitrary.  Under this generalization, the necessity of new gauge boson exchange in the s-channel could be relaxed for different reasons as we illustrate in the two concrete examples below.

\subsection{Unitarity in $\bg{SU(5)}$ gauge theory}

Among the rank 4 gauge groups containing the SM, $SU(5)$ is popular and has been well studied as a unified gauge symmetry.  Because of baryon number violation that occurs in this theory, the new gauge bosons of $SU(5)$, the $X$ and $Y$, must be very heavy, of order $10^{15}$ GeV or higher.  Nevertheless, it is of theoretical interest to see how unitarity is preserved in  processes such as $f \bar{f} \rightarrow X \bar{X}$.
Note that the general expressions we have derived based on unitarity would apply for such processes even though the $X$ and $Y$ gauge
bosons are colored and fractionally charged.  It is nontrivial to see the preservation of unitarity in the absence of a $Z'$ boson in this model.

The fermion-gauge boson interactions arise from the covariant derivative
$\mathcal{L}_{Af\bar f} = -g\bar{\psi}\gamma^\mu \bg{A}^T_\mu \psi + g~{\rm Tr}~\bar{\chi}\gamma^\mu (\bg{A}_\mu \chi + \chi \bg{A}^T_\mu)$,
where $\psi$ and $\chi$ stand for the left-handed fermion fields transforming as $\bg{5^*}$ and $\bg{10}$ respectively, and $\bg{A}_\mu \equiv A_\mu^iT^i$, $T^i$ are the $SU(5)$ generators, with ${\rm Tr}~(T^iT^j)=\frac{1}{2}\delta^{ij}$.
More explicitly, the above interaction can be written as
\begin{eqnarray}
\mathcal{L}_{Af\bar f} &=& \frac{g}{\sqrt{2}} \left[ \left( \bar{\nu} \gamma^\mu e + \bar{u}_\alpha \gamma^\mu d_\alpha \right) W^+_\mu -  \hat{X}^a_{\mu\alpha} \epsilon^{\alpha\beta\gamma} \bar{u}^c_\gamma \gamma^\mu Q_{\beta a} + \hat{X}^a_{\mu\alpha} \epsilon^{ab} \left(\bar{Q}_{\alpha b} \gamma^\mu e^c - \bar{L}_b \gamma^\mu d^c_\alpha \right)  + {\rm h.c.} \right] \nonumber \\
&& +~ \frac{g}{\cos\theta_W}\sum_f (T_3(f)-Q_f\sin^2\theta_W) \bar{f} \gamma^\mu f~Z_\mu +  e\sum_f Q_f\bar{f}\gamma^\mu f A_\mu \ ,
\label{fgsu5}
\end{eqnarray}
with $\tan\theta_W=\sqrt{{3}/{5}}$. Here $\alpha,\beta,\gamma$ are color indices, $\mu,\nu$ are Lorentz indices, and $a,b$ are $SU(2)$ indices. The new gauge bosons $\hat{X}_\alpha$ transform as $(3,2,-5/6)$ under $SU(3)_c \times SU(2)_L \times U(1)_Y$, which is written as
$\hat{X}_\alpha = (X^{-4/3}_\alpha,Y^{-1/3}_\alpha)$.
From the gauge kinetic term
$\mathcal{L}_{\rm kin} = -{\rm Tr} (F_{\mu\nu}F^{\mu\nu})/2$, one gets the trilinear gauge interactions:
\begin{eqnarray}
\mathcal{L}_{\rm gauge}^{\rm trilinear} &=& ig\cos\theta_W \hat{\mathcal{L}}_{W^+W^-Z} ~-~ ig\cos\theta_W \hat{\mathcal{L}}_{Y^*YZ}\! ~+~ \! ie \hat{\mathcal{L}}_{W^+W^-A}\! ~+~ \! i\frac{4}{3}e \hat{\mathcal{L}}_{X^*XA}\! ~+~ \! i\frac{e}{3}\hat{\mathcal{L}}_{Y^*YA}\! \nonumber \\
&& + ~\! i\frac{g}{\sqrt{2}} \hat{\mathcal{L}}_{Y^*XW^+}\! ~-~ \!i \frac{g}{\sqrt{2}}\hat{\mathcal{L}}_{X^*YW^-},
\end{eqnarray}
where $\hat{\mathcal{L}}_{A_1 A_2 A_3}$ are defined in Eq.~(\ref{gauge-trilinear}).
The implementations of the sum rules of Eq.~(\ref{sumrule1}) for different processes in the $SU(5)$ model are summarized in Table~\ref{su5}.
For the scattering process $f_1\bar{f}_2 \to A_1 A_2$ with $A_{1,2}=X,Y$,
one has to pay special attention to color contractions for each diagram. Take for example the $\nu\bar{\nu} \to YY^*$. Here, the scattering occurs through u-channel anti-down-quark exchange and s-channel $Z$ exchange. The incoming particle can only be left-handed. The color weights are the same for both diagrams. The sum rule Eq. (\ref{sumrule1}) implies $(g^L_{Y\nu d})^2 = -g^L_{Z\nu\nu}g_{Y^*YZ}$. Since $(g^L_{Y\nu d})=g/\sqrt{2}$ while $g_{Y^*YZ}=-g\cos\theta_W$ and $g^L_{Z\nu\nu}=g/(2\cos\theta_W)$, this sum rule is satisfied. A nontrivial color weight arises in the $u\bar{u} \to XX^*$ process which  occurs through t-channel anti-up quark exchange and s-channel photon exchange. Here both left-handed and right-handed initial quarks contribute. Since there is a color factor $\epsilon_{\alpha\beta\gamma}$ in the $uuX$ vertex, there will be 6 possible color contractions for the t-channel while the s-channel will have 9 contractions. The sum rule, then, reads (for initial left-handed $u$) $6(g/\sqrt{2})^2 = 9(Q_uQ_{X^*})g^2\sin^2\theta_W$. Since $SU(5)$ gauge theory predicts $\sin^2\theta_W = 3/8$, again we find that the above sum rule is satisfied. Other processes are summarized in Table \ref{su5}. We see that in this theory perturbative unitarity is preserved without the need for a $Z'$ boson.

\begin{table}[t!]
\centerline{\begin{tabular}{|c|l|l|c|} \hline
Process & Exchanged particle & Couplings & Color factor \\ \hline \hline
\raisebox{-2.2ex}[0pt]{$\nu \bar{\nu} \to Y Y^*$}  & t-channel anti-down quark \hspace{1.5cm} & $g^L_{Y\nu d}=\frac{g}{\sqrt{2}}$, ~$g^R_{Y\nu d}=0$ & 3\\
& s-channel $Z$ & $g_{Y^*YZ} = -g\cos\theta_W$ & 3 \\ \hline
\raisebox{-2.2ex}[0pt]{$u \bar{u} \to X X^*$}  & t-channel anti-up quark \hspace{1.5cm} & $g^L_{X^*uu}=-g^R_{X^*uu}=\frac{g}{\sqrt{2}}$ & 6 \\
& s-channel $\gamma$ & $g_{X^*X\gamma} = 2g^L_{\gamma u\bar{u}} = 2g^R_{\gamma u \bar{u}} = \frac{4}{3}e$ & 9 \\ \hline
 & t-channel anti-up quark & $g^L_{Y^*du} = -\frac{g}{\sqrt{2}}$, ~$g^R_{Y^*du}=0$ & 6 \\
$d\bar{d} \to YY^*$	 & u-channel anti-neutrino &  $g^R_{Y d\nu} = \frac{g}{\sqrt{2}}$, ~$g^L_{Y d\nu}=0$ & 3 \\
& s-channel $\gamma,Z$ & $g_{Y^*YA} = \frac{e}{3}$, ~$g_{Y^*YZ} = -g\cos\theta_W$ & 9 \\ \hline
 & t-channel anti-up quark & $g^L_{X^*uu}=g^L_{Y^*du}=-\frac{g}{\sqrt{2}}$, ~$g^R_{Ydu}=0$ & 6 \\
$u\bar{d} \to X^*Y$ & u-channel positron & $g^L_{Xde} = -g^L_{Y^*ue} = \frac{g}{\sqrt{2}}$, ~$g^R_{Yue} = 0$ & 3 \\
 & s-channel $W^+$ & $g_{Y^*XW^+} = \frac{g}{\sqrt{2}}$ & 9 \\ \hline
\end{tabular}}
\caption{Summary of $f_1\bar{f}_2 \to A_1 A_2$ with $A_{1,2}=X,Y$ processes in the $SU(5)$ theory. One sees that the sum rules -- after accounting for  the color factors -- given by Eqs. (\ref{sumrule1}) are satisfied.}
\label{su5}
\end{table}

\subsection{Unitarity in $\bg{SU(3)_c \times SU(2)_L \times SU(2)'}$ model}

It is possible to embed the hypercharge $Y$ of the SM consistently into an $SU(2)$ symmetry, if one allows for exotic fermions.
A consistent theory emerges with the following fermionic assignment under $SU(3)_c \times SU(2)_L \times SU(2)'$.
\begin{equation}
Q_L(3,2,2), \quad Q_L'(3,1,1), \quad  Q_R(3,1,5), \quad  \psi_L(1,2,4), \quad \psi_R(1,1,7) \ .
\end{equation}
Hypercharge is obtained as $Y/2 = T_3'/3$ in this model.
The field $Q_L$ contains left-handed quark doublet and two exotic quarks with charges opposite to the SM quarks: $u'_L~(Q=-2/3, ~T_3=-1/2)$ and $d'_L~(Q=1/3, ~T_3=1/2)$. The $Q_R$ field contains the SM right-handed quarks $u_R$ and $d_R$.  In addition it has three exotic quarks:
 $u'_R~(Q=-2/3)$, $d'_R~(Q=1/3)$, and a neutral quark $q^0_R$. The left-handed and right-handed leptons are contained in the $\psi_L$ and $\psi_R$ respectively. Here we focus on the quark sector. The quark--gauge boson interactions are given by
\begin{eqnarray}
\mathcal{L}_{ff{\rm gauge}} = {\rm Tr}~ \bar{Q}_L \gamma^\mu (g \bg{A}_\mu Q_L + g' Q_L \bg{A}'_\mu) ~+~ g'\bar{Q}_R \gamma^\mu \bg{A}'_\mu Q_R\,.
\end{eqnarray}
The identification of hypercharge leads to $g'=g_Y/3$ with $g_Y=g\tan\theta_W$. Explicitly we have
\begin{eqnarray}
\mathcal{L}_{ff{\rm gauge}} &=& \left[\frac{g}{\sqrt{2}} \left(\bar{u}_L \gamma^\mu d_L + \bar{d}'_L \gamma^\mu  u'_L \right)W^+_\mu + \frac{g}{3\sqrt{2}}\tan\theta_W \left(\bar{u}_L \gamma^\mu d'_L + \bar{d}_L \gamma^\mu  u'_L \right)W'^{+}_\mu \right.  \nonumber \\
&& \left. + \frac{g\sqrt{2}}{3}\tan\theta_W \left(\bar{u}_R \gamma^\mu d'_R + \bar{d}_R \gamma^\mu  u'_R \right) W'^+_\mu  + \frac{g}{\sqrt{3}}\left(\bar{d}'_R\gamma^\mu q^0_R + \bar{q}^0_R \gamma^\mu d_R\right)W'^+_\mu + {\rm h.c.} \right] \nonumber \\
&& +~ \mathcal{L}_{\gamma ff} + \mathcal{L}_{Zff},
\label{su2plag}
\end{eqnarray}
where $\mathcal{L}_{\gamma ff}$ and $\mathcal{L}_{Zff}$ have the same form as in the SM. In this model the $W'^+$ gauge boson carries $+{1}/{3}$ electric charge.  The trilinear gauge interactions are given by
\begin{eqnarray}
\mathcal{L}_{\rm trilinear} &=& i\frac{e}{3}\hat{\mathcal{L}}_{W'^+W'^-A} - i\frac{e}{3}\tan\theta_W \hat{\mathcal{L}}_{W'^+W'^-Z} + ig \cos\theta_W \hat{\mathcal{L}}_{W^+W^-A} + ie \hat{\mathcal{L}}_{W^+W^-Z} \ ,
\end{eqnarray}
where $\hat{\mathcal{L}}_{A_1A_2A_3}$ is in the form of Eq. (\ref{gauge-trilinear}).

In order to see how unitarity is preserved here, we study $u\bar{u} \to W'^+W'^-$ scattering. This process  occurs through t-channel $d'$ exchange, s-channel photon exchange, and s-channel $Z$ exchange. If the incoming $u$ quark is left-handed (in the massless limit), the sum rule Eq. (\ref{sumrule1}) reads $(g^{L}_{W'^+u\bar d'})^2 = g^2(Q_uQ_{W'^+})\sin^2\theta_W -g^2\tan^2\theta_W\left(1/2-Q_u\sin^2\theta_W\right)/3 $, which is
satisfied since $g^L_{W'^+u\bar d'} = g\tan\theta_W/(3\sqrt{2})$. If the incoming $u$-quark is right-handed, the sum rule gives $(g^R_{W'^+u\bar d'})^2 = g^2(Q_uQ_{W'^+})\sin^2\theta_W + g^2\tan^2\theta_W\left(Q_u\sin^2\theta_W\right)$ which is satisfied since $g^R_{W'^+u\bar d'} = g\sqrt{2}\tan\theta_W/3$, from Eq. (\ref{su2plag}).

We summarize the $f_1 \bar{f}_2 \to A_1 A_2$ processes in Table \ref{su2prime}. In this model, the problem with infinite chain of fermions described in the beginning of Sec. \ref{genaralW'} is evaded because $W'$ boson here always couples to a SM fermion and an exotic fermion. Therefore a closed loop can be formed in a moose-like diagram.

\begin{table}[t!]
\centerline{\begin{tabular}{|l|l|l|} \hline
Process & Exchanged particle & Couplings  \\ \hline \hline
\raisebox{-2.2ex}[0pt]{$u \bar{u} \to W'^+W'^-$} \hspace{1.0cm} & t-channel $d'$ \hspace{1.5cm} & $g^R_{W'^+u\bar{d}'} = 2g^L_{W'^+u\bar{d}'}=\frac{g\sqrt{2}}{3}\tan\theta_W$ \\
 & s-channel $\gamma,Z$ & $g_{W'^+W'^-Z} = -\frac{e}{3}\tan\theta_W$ \\ \hline
\raisebox{-2.2ex}[0pt]{$u \bar{u}' \to W^+W'^+$} \hspace{1.0cm} & t-channel $d$ \hspace{1.5cm} & $g^L_{W^+u\bar{d}} =  \frac{g}{\sqrt{2}}$, ~$g^L_{W'^+d\bar{u}'} = \frac{g}{3\sqrt{2}}\tan\theta_W$, ~$g^R_{W^+u\bar{d}} = 0$ \\
 & u-channel $d'$ & $g^L_{W^+d'\bar{u}'} =\frac{g}{\sqrt{2}}$, ~$g^L_{W'^+u\bar{d}'} = \frac{g}{3\sqrt{2}}\tan\theta_W$, ~$g^R_{W^+u'\bar{d}'} = 0$ \\ \hline
 & t-channel $u$ & $g^R_{W'^+u\bar{d}'} = 2g^L_{W'^+u\bar{d}'} = \frac{g\sqrt{2}}{3}\tan\theta_W$ \\
$u\bar{d}' \to Z W'^+$ & u-channel $d'$ & $g^R_{W'^+u\bar{d}'} = 2g^L_{W'^+u\bar{d}'} = \frac{g\sqrt{2}}{3}\tan\theta_W$ \\
 & s-channel $W'^+$ & $g_{Zd'\bar{d}'} = \frac{g}{\cos\theta_W}(\frac{1}{2}-\frac{1}{3}\sin^2\theta_W)$  \\ \hline
\end{tabular}}
\caption{Summary of $f_1\bar{f}_2 \to A_1 A_2$ processes in the $SU(3)_c \times SU(2)_L \times SU(2)'$ theory. One sees that the sum rules given by Eqs. (\ref{sumrule1}) are satisfied. In this model the $W'^+$ carries $+1/3$ electric charge.}
\label{su2prime}
\end{table}

\section{Conclusion}

In this paper, we have studied perturbative unitarity constraints on general $W'$ models arising from fermion scattering into final state
gauge bosons.  The requirement of unitarity enables us to derive one set of sum rules
for the couplings, and a second set of sum rules for the mass scales of various particles involved in the scattering process.
We have investigated implications of these constraints on hadron collider physics.
In most models, a $Z'$ gauge boson  is required to accompany the $W'$ in order to unitarize the theory.
Unitarity fixes uniquely all couplings of the $Z'$ to the SM fermions and gauge bosons. The finite sum rule enables us to derive upper limit on the
$Z'$ gauge boson mass, which we find typically to be not much above the $W'$ mass.
We applied these results to a class of $W'$ models that have been proposed to explain the anomalous $t\bar t$ asymmetry reported by the Tevatron
experimnents.  In these models, the $Z'$ mass is found to lie below $(7-8)$\,TeV. The allowed $Z'$ mass window has already been tightly constrained by hadron collider experiments below a TeV. The LHC can probe higher mass regions all the way up to the unitarity limit and could thus falsify this class of models. We have observed an interesting role unitarity may play in the context of $W'$ models in explaining the $Wjj$ event excess reported by the CDF collaboration. By making the $Z'$ relatively heavy, but consistent with the unitarity limit, it is possible to generate the needed
cross section for $p \overline{p} \rightarrow W^\pm W'^{\mp}$ with the $W'^{\mp}$ decaying into two jets, while suppressing new contributions
to the $Zjj$ cross section.  In the last section, we have developed models with an extended gauge symmetry that has rank 4, thus having extra gauge boson, but not a neutral $Z'$. We have classified the finite set of possibilities in this class.
By studying two concrete examples, we have shown how unitarity is preserved in this class of models without a $Z'$.
Hopefully the model--independent constraints derived here from unitatiry would find use in $W'$ model building and its experimental tests.

\section*{Acknowledgement}

We thank M.-C.~Chen, M.~Nemev\v{s}ek, F.~Nesti, S.~K.~Rai, G.~Senjanovi\'c, A. Yu.~Smirnov and H.-B.~Yu for useful discussions.
The work of KSB and JJ is supported by the US Department of Energy, Grant Numbers DE-FG02-04ER41306.
The work of YZ is supported in part by the National Science Foundation under Grant No. 1066293 via the Aspen Center for Physics.
YZ would like to thank the Department of Physics, Oklahoma State University for hospitality where part of the work was done.

\end{document}